\documentclass[letterpaper, 10 pt, conference]{ieeeconf}  

\IEEEoverridecommandlockouts                              

\usepackage{amsmath,amssymb,amsfonts}
\usepackage{algorithmic}
\usepackage{graphicx}
\usepackage{subcaption}
\usepackage{textcomp}
\usepackage{wrapfig,lipsum,booktabs}
\usepackage{booktabs}
\usepackage{caption}
\usepackage{wrapfig}
\usepackage[utf8]{inputenc}

\usepackage{graphicx}
\usepackage{mathrsfs}
\usepackage{amsfonts}
\usepackage{amsmath}
\usepackage{amssymb}
\usepackage{lipsum}
\usepackage{url}
\usepackage{subfloat}
\newcommand{\argminF}{\mathop{\mathrm{argmin}}\limits}   
\DeclareMathOperator*{\argmin}{arg\,min}

\begin{document}

\title{Unraveling the Control Engineer's Craft with Neural Networks} 
		
\author{Braghadeesh Lakshminarayanan, Federico Dettù, Cristian R. Rojas, Simone Formentin \thanks{F. D. and S. F. are with the Dipartimento di Elettronica, Informazione e Bioingegneria, Politecnico di Milano, 20133, Milano, Italy. Emails: federico.dettu@polimi.it (F. D.), simone.formentin@polimi.it (S. F.).}\thanks{B. L. and C. R. are with the Division of Decision and Control Systems, KTH Royal Institute of Technology, 100 44, Stockholm, Sweden. Emails:  blak@kth.se (B. L.), crro@kth.se (C. R.).}
}
		

\maketitle

\begin{abstract}                
Many industrial processes require suitable controllers to meet their performance requirements. More often, a sophisticated digital twin is available, which is a highly complex model that is a virtual representation of a given physical process, whose parameters may not be properly tuned to capture the variations in the physical process. In this paper, we present a \textit{sim2real}, direct data-driven controller tuning approach, where the digital twin is used to generate input-output data and suitable controllers for several perturbations in its parameters. State-of-the art neural-network architectures are then used to learn the controller tuning rule that maps input-output data onto the controller parameters, based on artificially generated data from perturbed versions of the digital twin. In this way, as far as we are aware, we tackle for the first time the problem of re-calibrating the controller by \emph{meta-learning the tuning rule} directly from data, thus practically \textit{replacing the control engineer with a machine learning model}. The benefits of this methodology are illustrated via numerical simulations for several choices of neural-network architectures. 
\end{abstract}
	
\section{Introduction}
\label{Section:Introduction}
Tuning a control system is not an easy task. Several factors need to be taken into account, and semi-empirical or heuristic rules are often followed by practitioners -- see, \emph{e.g.}, the Ziegler-Nichols or Cohen-Coon methods~\cite{aastrom2006advanced}. Furthermore, given that many control systems are required to operate in varying conditions, special care to robustness or adaptation is necessary, which often turns to the necessity of re-calibrating it.
This constitutes one of the reasons behind the known detachment between control theorists and practitioners, as discussed in the recent survey~\cite{samad2017survey}. \\
Being the rules used in controller tuning often heuristic, they are not easily encoded in formal algorithms, and they might require a significant amount of work by the field engineer manipulating the regulator tuning knobs -- this is also known as \emph{end-of-line controller tuning} \cite{tanelli2011transmission,dettu2023til}.\\ 
The recent spread of high-fidelity in-silico replica of real plants, denoted as \textit{digital twins} (DT) \cite{liu2021review}, opens up further possibilities; in fact, DTs can be employed to generate substantial data volumes cost-effectively, possibly in a wide amount of scenarios, \emph{e.g.}, by modifying some typically uncertain system parameters. \\
However, DTs are also used for designing and testing controllers \cite{dettu2023til}, due to the possibility of executing cheap tests and without any kind of safety hazard. Typically, one could modify some system parameters and then find via the use of a DT the best set of control parameters for this occurrence.
As a consequence, a DT can be in principle used to generate 1) a large set of data and 2) tune a controller for it, given some design rule. In this research, we employ such amount of information to learn from data the tuning rule, approximating it via some class of functions. In a sense, we \textit{meta-learn} the control design procedure.\\ \\
Many approaches exist for solving the generic problem of \textit{extracting} or \textit{learning} from data the controller parameters: the classical solution to this problem is the identification of the plant and subsequent application of the control tuning rule~\cite{ljung1999system}. However, more efficient solutions have been proposed, which aim at directly estimating the control parameters from data without the need for a model \cite{piga2017direct,formentin2012non,formentin2014comparison} -- these methods attempt at minimizing the difference between the closed-loop performance of the system under control and some ideal reference model. Other approaches rely on black-box optimizers, such as Particle Swarm \cite{qi2019tuning} or Bayesian Optimization (BO) \cite{khosravi2021performance,lucchini2020torque}: they iteratively minimize a performance metric, with the big advantage that no analytical form for such function is necessary, due to the approaches being purely black-box -- \emph{e.g.}, BO constructs from data a Gaussian Process surrogate of the cost function.\\ \\
Nonetheless, the approaches above do not exploit the huge amount of data enabled by high-fidelity DTs; also, such approaches usually constrain in some way the controller structure or the given plant, while not directly addressing the problem of \textit{learning} how to design the controller itself. For example, \cite{piga2017direct} addresses Linear Parameter Varying systems, and requires the use of a Predictive Controller as an outer governor. \cite{formentin2012non} is specifically tailored for linear systems, and learns linear controllers. \\ In addition, if a practitioner is using some empirical tuning rule, which is however hardly numerically encoded, but still satisfactory from a performance point of view, the methods above are of no use in trying to replicate the same tuning rule.
\\
The contributions of this article are twofold. We propose a method to directly exploit the availability of large amounts of data to simplify the design of a controller, via black-box function; this approach results in a direct mapping from data to control parameters, which allows for potentially learning the control design procedure per-se, and does not impose any particular constraint on the controller structure, nor on the system under control.
\\
The rest of the paper is structured as follows: In Section \ref{Section:Problem_statement}, we formally state the addressed problem, while we propose the learning of the control design procedure in Section \ref{section:control_design}. Then, in Section \ref{section:case_study} we define a suitable case study -- the yaw-rate tracking problem -- on which we apply the proposed learning approach in Section~\ref{section:simulation_analysis}. Finally, conclusions are drawn in Section~\ref{sec:conclusion}.

\section{Problem statement}
\label{Section:Problem_statement}
In the following, we describe the setup we consider in this work, and define the problem we aim at solving.
Consider a single-input single-output discrete-time system $\mathcal{M}(\Theta)$
\begin{equation}
	\mathcal{M}: \begin{cases}
		x_{k+1}=f(x_k,u_k,\Theta), \\
		y_k = g(x_k, \Theta),
	\end{cases}
\label{eq:system_eq}
\end{equation}
where $x_k\in \mathbb{R}^{n_x}$, $u_k\in \mathbb{R}$ and $y_k\in \mathbb{R}$ are the state vector, input and output of the system. $f$ and $g$ are state and output functions, possibly non-linear. $\Theta \in \mathbb{R}^{p}$ is a vector of uncertain parameters, $\Theta=\left[\theta_1,\ldots,\theta_p\right]$, each distributed according to a probability distribution $\theta_i \sim \mathcal{D}_i$. 
We assume that we are able to obtain input-output sequences, say, $D_1^{N},\ldots, D_M^{N}$, of length $N$, from an available simulator of $\mathcal{M}$, given a specific value of the parameter vector. \\
Further, consider a parametric controller $C(\phi)$, where  $\phi\in \mathbb{R}^{n_\phi}$ is the vector of controller parameters, and assume that a control design rule $\mathcal{R}\colon \mathcal{M}\left(\Theta\right) \mapsto \phi$ is available. \\
The assumptions above are realistic: if the simulator is a white-box, this means that we have a somewhat high knowledge of the underlying model. We might however have no real knowledge of the plant, and this happens when the simulator is a black-box (for example, this happens when dealing with commercial driving simulators \cite{dettu2023til}), and the user can only set some parameters and look at the model responses. \\    
Under these considerations, we are able to collect a set of data, depicted in Table \ref{tab:table_parameters}. 

\begin{table}[h]
\centering
	\begin{tabular}{|l|l|l|}
		\hline
		$\Theta_1$ & $D_1^N=\left\{\left(u_1^{(1)},y_1^{(1)}\right),\ldots,\left(u_N^{(1)},y_N^{(1)}\right)\right\}$ & $C(\phi_1)$ \\ \hline
		$\Theta_2$ & $D_2^N=\left\{\left(u_1^{\left(2\right)},y_1^{\left(2\right)}\right),\ldots,\left(u_N^{\left(2\right)},y_N^{\left(2\right)}\right)\right\}$ & $C(\phi_2)$ \\ \hline
		$\vdots$   & $\vdots$                                                                     & $\vdots$    \\ \hline
		$\Theta_M$ & $D_M^N=\left\{\left(u_1^{\left(M\right)},y_1^{\left(M\right)}\right),\ldots,\left(u_N^{\left(M\right)},y_N^{\left(M\right)}\right)\right\}$ & $C(\phi_M)$ \\ \hline
	\end{tabular}
\caption{Collection of model parameters, input-output trajectories and designed controllers.}
\label{tab:table_parameters}

\end{table}
\vspace{-1em}
Each row of the table shows how a realization of the parameters $\Theta_j$, for $j=1,\ldots,M$, is mapped onto a collection of input-output time-domain data -- the pairs $(u_i^{\left(j\right)},y_i^{\left(j\right)}),\ i=1,\ldots,N$ -- and a parametric controller $C\left(\phi_j\right)$.\\
Given this framework, we are able to design a controller for a given realization of the system $\mathcal{M}(\Theta)$ which is optimal -- in the sense of satisfying the tuning rule $\mathcal{R}$. But what happens when $\Theta$ varies during the system operation? While a classical solution to this problem consists in re-identifying the system parameters, we here want to leverage the available data to find a \textit{direct} solution to get the controller parameters, \textit{de-facto} learning the tuning rule $\mathcal{R}$.

\section{Data-driven control design learning}
\label{section:control_design}
Given the available information in Table \ref{tab:table_parameters}, we want to find the best control parameters satisfying tuning rule $\mathcal{R}$. At run-time, a mere series of time-domain information is available. We can exploit these data so as to estimate the controller parameters directly, in a data-driven fashion. Precisely, we want to learn function $\mathcal{F}\colon D_i^{N} \mapsto \hat{\phi}$ as
\begin{equation}
	\mathcal{F}^*=\argminF_{\mathcal{F}} \dfrac{1}{M}\sum_{i=1}^{M}\left\|\phi_i-\mathcal{F}\left(D_i^{N}\right)\right\|^2,
	\label{eq:function_bb}
\end{equation}
which implies that we are seeking for the best estimator of the controller parameters given the available data. In general, finding $\mathcal{F}$ in a general setting is not an easy task, as this means extracting some meaningful features from time-domain data, and mapping these features -- that have to intrinsically contain some information about the uncertain parameters -- onto the best controller parameters. The problem thus translates into the selection of a proper class of functions, for which neural networks (NNs) are suitable candidates. In this paper, we consider  Convolutional Neural Networks (CNNs) and WaveNets (WNs) to solve Eq.~\eqref{eq:function_bb}.\\

\textit{A. CNN-based controller learning}

\begin{figure*}[t!]
\vspace{1em}
	\centering
	\includegraphics[width=1.5\columnwidth]{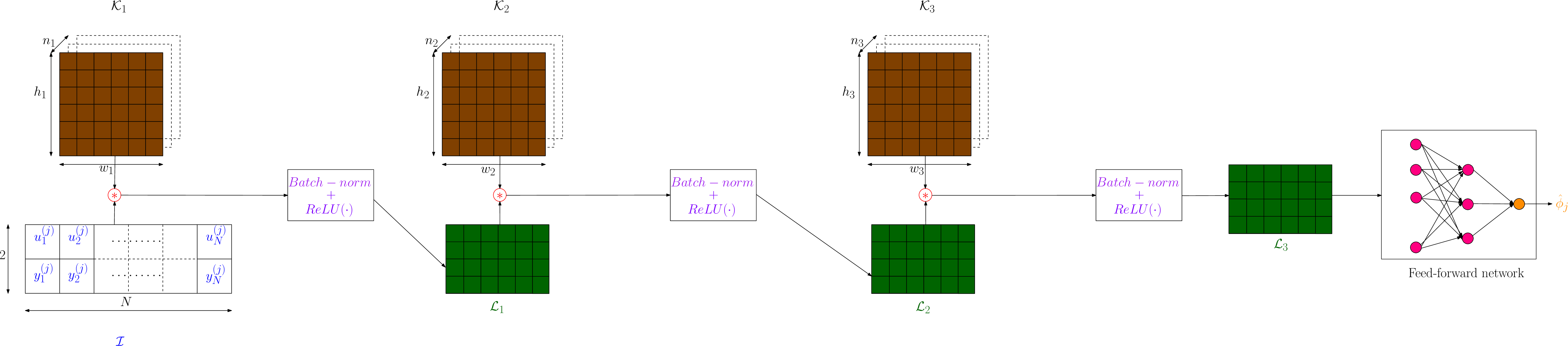}
	\caption{Architecture of the considered CNN. $\mathcal{K}_i$ denotes a filter bank that consists of kernels of dimensions $h_i \times w_i \times n_i$, followed by batch-norm and ReLU operations. $h_i$ and $w_i$ denote the height and width of a kernel respectively and $n_i$ denotes the number of filters. $\mathcal{L}_i$ is the output corresponding to $\mathcal{K}_i$.}
	\label{fig:cnn_architecture}
\end{figure*}

CNNs are a particular class of non-linear functions with an extremely high approximation capability; in fact, they have been used for a long time in image classification tasks, but applications with time-domain data recently emerged \cite{linaro2023continuous,karlsson2021speed}. The considered CNN architecture is displayed in Fig.~\ref{fig:cnn_architecture}. The model input is $\mathcal{I}\in \mathbb{R}^{2 \times N}$: in practice, system input and output time-series data are collected row-wise. Each column of $\mathcal{I}$ represents the $k$-th time step for the input-output pair.
Three series of 2-D convolution filters are applied to $\mathcal{I}$; a $ReLU$ transformation is applied after each bank of filters\footnote{The ReLU function is defined as $ReLU(x) := \max \left(0,x\right)$.}. Details on this kind of neural networks are given in \cite{wu2017introduction}.
The last layer is a Fully Connected one, and provides the estimate of the controller parameters $\hat{\phi}$.
The number of filters and kernel sizes in the actual implementation are discussed later on, in Section \ref{section:simulation_analysis}.
Indeed, for training purposes, other layers are added, following common practice in CNN design, namely batch normalization, dropout and regression ones. 

\medskip

\textit{B. WN-based controller design}

\begin{figure}[h]
\centering
\begin{subfigure}{0.3\textwidth}
    \centering \includegraphics[width=1.1\columnwidth]{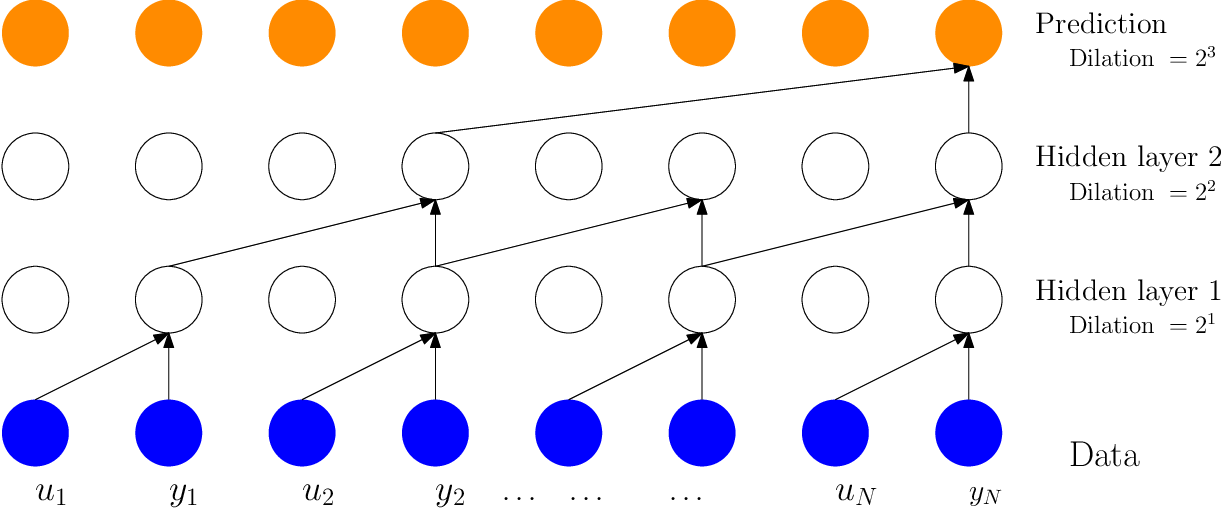}
    \caption{WaveNet architecture for prediction.}
    \label{fig:Wavenet_a}
    \end{subfigure}

    \medskip
    \begin{subfigure}{0.3\textwidth}
    \centering \includegraphics[width=1.1\columnwidth]{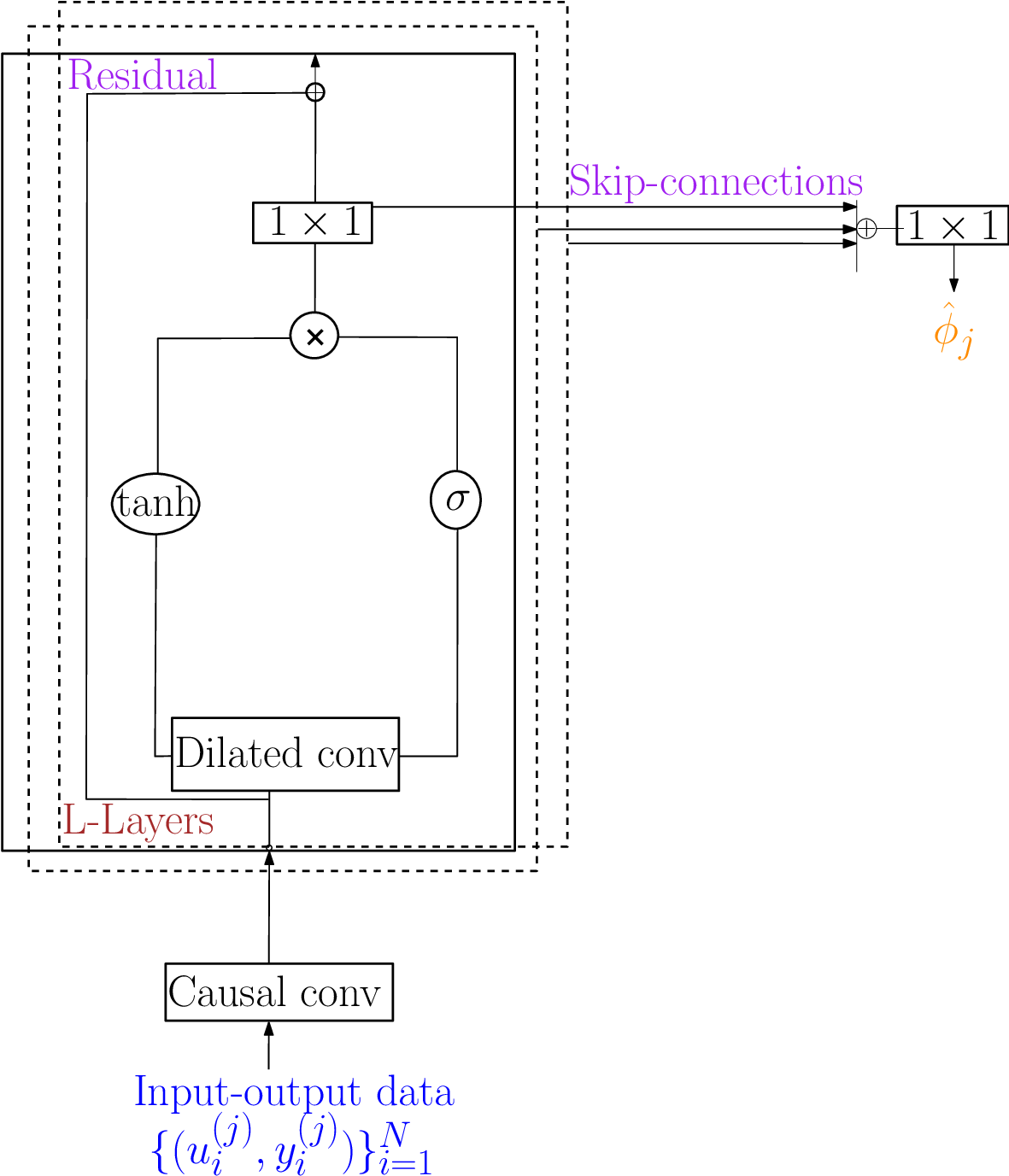}
    \caption{Operations performed in the WaveNet.}
    \label{fig:Wavenet_unit}
    \end{subfigure}
\caption{WaveNet architecture and operations.}
\label{Fig: Wavenet}
\end{figure}

WNs~\cite{oord2016wavenet} are auto-regressive neural-network models that capture long term dependencies in the data. These models were originally proposed for audio signal processing to predict the next word in the audio or translate text to speech. A WN model is based on the principle of causal 1-D convolution: The outputs of a given hidden or output layer depend only on the present and past values of the outputs of the preceding layer. The range of such dependencies is pre-specified by a kernel of size $K$, which performs the convolution. This is different from the vanilla CNN where the outputs have non-causal dependencies. To capture the long-term dependencies, dilations are introduced of order of $2^l$, where $l=0,1,\ldots,L-1$ and $L$ is the number of layers. A schematic of the WaveNet architecture is shown in Fig.~\ref{fig:Wavenet_a}.

The input layer of a WN architecture has input-output data from a dynamical system arranged as an alternating sequence, denoted as $\{(u_i,y_i)\}_{i=1}^N$, where $u_i$ represents the input at time $i$ and $y_i$ is the corresponding output. 

Each neuron in a hidden layer of a WN is an activation function that is composed of several mathematical operations as shown in Fig.~\ref{fig:Wavenet_unit}. During the training phase of a WN, the causal 1-D convolution is performed on each input-output data sequence $(u_i^{(j)},y_i^{(j)})$, whose corresponding controller parameters are $\phi_j$ (see Table~\ref{tab:table_parameters}).
The outputs of the convolution are then fed to  the $L$ stacked hidden layers as shown in Fig.~\ref{fig:Wavenet_unit}.

Each hidden layer consists of a dilated convolution followed by two parallel operations: (i) the outputs of the dilated convolution are first mapped by \emph{tanh} and \emph{sigmoid ($\sigma$)} operations, and then multiplied, followed by a 1-D convolution (denoted by the $1 \times 1$ box in Fig.~\ref{fig:Wavenet_unit}), and (ii) a residual operation, where the outputs of the causal convolution are added with the outcomes of the $1 \times 1$ block corresponding to (i). The first operation is known as a \emph{skip connection}. The output layer of the WN architecture is the sum of skip-connections of $L$ stacked hidden layers followed by a 1-D convolution. The input to the WN has length $2N$ (${(u_i^{(j)},y_i^{(j)})}$), whereas its output is an another sequence of length $2N$, which is aggregated to yield $\hat{\phi}_j$ (the estimate of the controller parameters $\phi_j$). For example, the last entry of the sequence is taken as the estimate of $\phi_j$. 

By choosing appropriate values for $K$ and $L$, the estimate of $\phi_j$ can capture the long-term dynamics of time domain input-output data $(u_i^{(j)},y_i^{(j)})$, $i=1,\ldots,N$, $j=1,\ldots,M$.

\section{Case study: the yaw-rate tracking problem}
\label{section:case_study}
As a case study to prove the methodology described in Section \ref{section:control_design}, we consider the yaw-rate tracking problem, which is of high relevance within the automotive context.

\begin{figure}[h]
    	\centering
	\includegraphics[width=0.5\columnwidth]{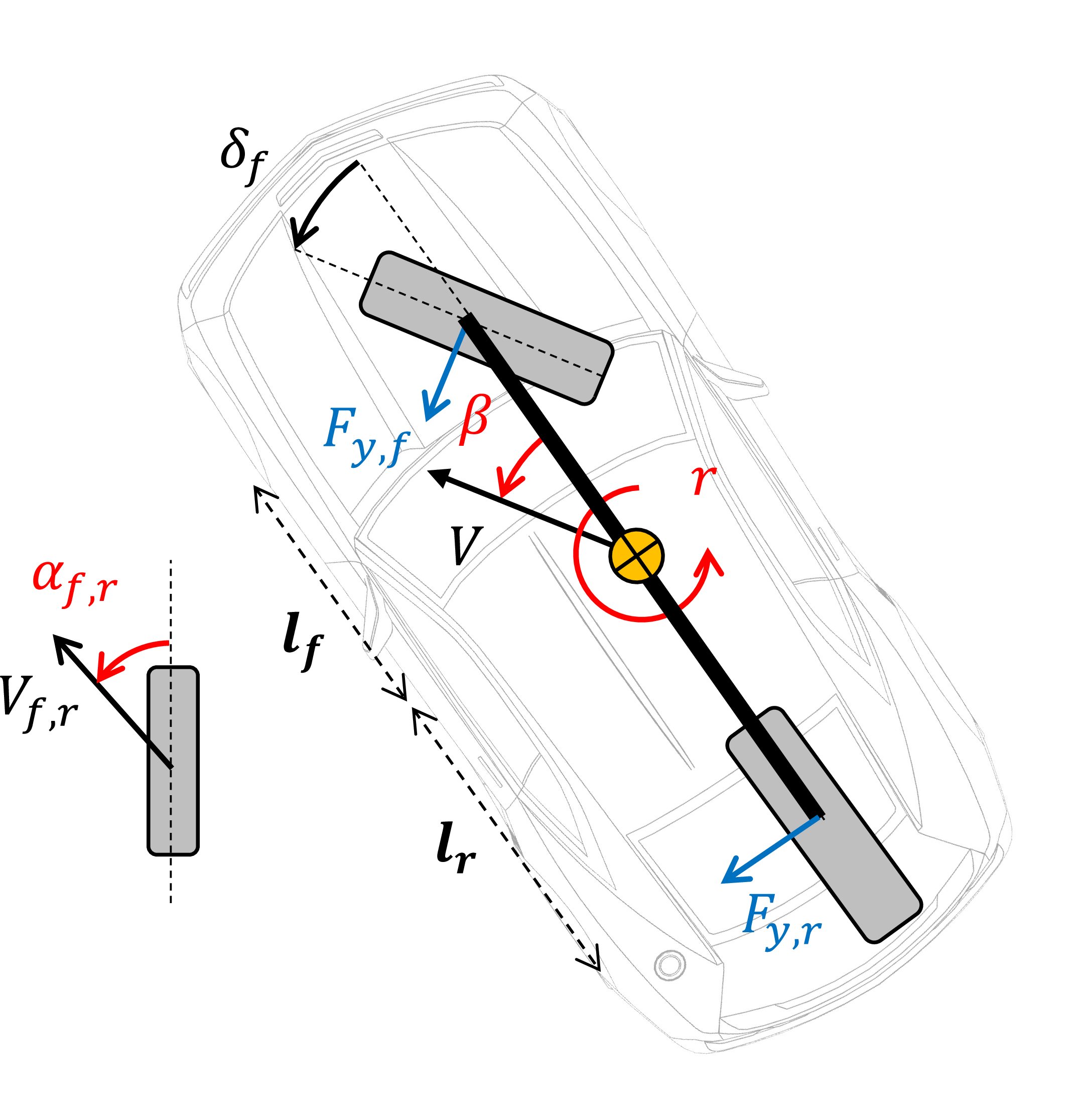}
	\caption{Schematic representation of the single-track model.}
	\label{fig:bicycle_model}
\end{figure}

\textit{1) System modelling:} Upon the assumptions of small steering angle $\delta_f$ and front/rear wheels side-slip angles $\alpha_{f,r}$, the following set of equations describes the yaw-rate dynamics:
\begin{equation}
	\begin{split}
		x_{k+1}=& Ax_{k}+Bu_k, \\
		y_k=&Cx_{k}.
	\end{split}
	\label{eq:system_bicycle_model}
\end{equation}
The model state is $x_k=\begin{bmatrix} \beta_k & r_k \end{bmatrix}^T$, the input is $u_k=\delta_{f,j}$, and the output is $y_k=r_k$.
$\beta_k$ and $r_k$ represent the vehicle side-slip angle and yaw-rate, respectively (both depicted in Fig.~\ref{fig:bicycle_model}). Matrices $A$, $B$ and $C$ are given by
\begin{align}
	A=&T_s\begin{bmatrix}
	T_s^{-1}-\dfrac{C_f+C_r}{M_{veh}v_x} & \dfrac{-L_f C_f+L_r C_r-M_{veh}v^2_x}{M_{veh}v^2_x}\\
	\dfrac{-L_f C_f+L_r C_r}{J_{z}} & T_s^{-1}-\dfrac{L_f^2 C_f+L_r^2 C_r}{J_{z}v_x}
\end{bmatrix}, \nonumber \\
&B=T_s \begin{bmatrix}
	\dfrac{C_f}{M_{veh} v_x}  & \dfrac{C_f l_f}{J_z}
\end{bmatrix}^T, C=\begin{bmatrix}
0 &1
\end{bmatrix}.
\label{eq:matrices_system}
\end{align}

$T_s$ is the discretization time step, set here as $0.02$ s (\emph{e.g.}, as in \cite{lucchini2020torque}). Concerning the physical parameters appearing in the matrices, $M_{veh}$ is the vehicle mass, $J_z$ its inertia, and $l_{f,r}$ the distances between the center-of-gravity and the front-rear wheel axles, respectively. Then, $v_x$ is the vehicle longitudinal speed: this is not a physical parameter, but rather a time-varying state of the system; however, it is common practice to consider it as a constant in the yaw-rate control problem \cite{lucchini2020torque,beal2012model}. \\
The lateral forces in the model in Eq. \eqref{eq:system_bicycle_model} are linearly related to the wheel slip angles $F_y=-C_{\alpha} \alpha$ \cite{beal2012model}. The cornering stiffness coefficient is however highly dependent on the type of terrain, the tire characteristics, temperature and so on \cite{sierra2006cornering}. For this reason, it is to be considered an uncertain parameter, prone to significant variation throughout the vehicle functioning:
\begin{equation}
	F_{y}=-C_{\alpha} \alpha \cdot \mu_s,
\end{equation}
where $\mu_s$ is considered as a scaling coefficient for the tire-road friction. \\ Another typical source of uncertainty is the vehicle mass, which usually changes about some nominal value, typically including the chassis and tire payload, as well as the driver one: $M_{veh} = M_0 + M_{\delta}$, 
where $M_{\delta}$ is the mass variation.\\
For the considered model, we have identified two typical sources of uncertainty, the mass and the tire-road friction, which can be collected in the vector $\Theta \in \mathbb{R}^{2}$, and are both assumed to be \emph{a priori} uniformly distributed:
\begin{equation}
	\Theta=\begin{bmatrix}\mu_s & M_{\delta}	\end{bmatrix},\ \mu_s\sim \mathcal{U}\left(0.3,1.1\right),\ M_{\delta}\sim \mathcal{U}\left(0,300\right).
\end{equation}
The values of the model parameters are given in Table~\ref{tab:bicycle_params}.

\begin{table}[h]
\centering
\scalebox{0.8}{
\begin{tabular}{@{}cccc@{}}
\toprule
$M_{veh}$ {[}kg{]}                             & $l_{f,r}$ {[}m{]} & $J_z$ {[}kg $\cdot$ m$^2${]} & $C_{f,r}$ {[}N/(rad $\times 10^3$){]} \\ \midrule
$1729$                                           & $1.49,\ 1.16$       & $2483$                    & $1.42,\ 2.93$             \\
\midrule
\end{tabular}
}
\caption{Vehicle model physical parameters (for a two-seats sports car).}
\label{tab:bicycle_params}
\end{table}

\begin{figure}
   	\centering
	\includegraphics[width=0.9 \columnwidth]{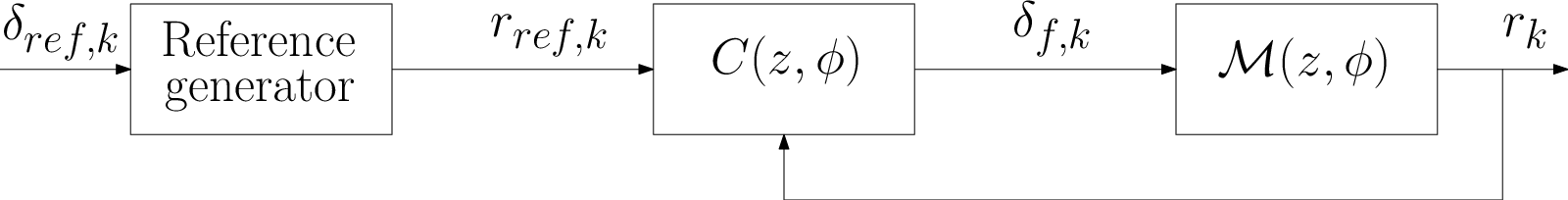}
	\caption{Control scheme for the yaw-rate tracking problem.}
	\label{fig:control_scheme}
\end{figure}
\textit{2) Control design:} The goal in yaw-rate tracking problems is that of regulating reference behaviour, generated by some ad-hoc logic based on the driver steer request; in practice, the driver intent is tracked electronically, and this helps avoiding unsafe situations -- \emph{e.g.}, oversteering, which might destabilize the vehicle. Overall, the control scheme is schematically represented in Fig.~\ref{fig:control_scheme}: the reference generator block converts the driver demand to yaw-rate requests, and many examples can be found in the literature  \cite{beal2012model,lucchini2020torque}.\\
One typical approach for designing controllers for linear time-invariant systems is loop shaping (used, \emph{e.g.}, for the same system in \cite{corno2021optimization}), which can be used to guarantee some phase margin $\phi_m$ and cutting frequency $\omega_c$. The latter two parameters then define the closed-loop behaviour of the system, according to well known basic control theory. A Proportional-Integral controller suffices for the system in \eqref{eq:system_bicycle_model}:
\begin{equation}
	C\left(z,\phi\right)=\dfrac{k_p}{1+2T_i/T_s}\dfrac{z+1}{z+\dfrac{T_s-2T_i}{T_s+2T_i}},
\end{equation}
where the controller is discretized via the Tustin approach (at sampling time $T_s$), and the tuning knobs are constituted by proportional gain and integral time $\phi=\begin{bmatrix} k_p & T_i \end{bmatrix}$.\\
By properly selecting the control parameters, one can achieve the desired frequency-domain performance. We require $\varphi_m \geq 80^{\circ}$ -- a basic value to have a well-damped first-order system-- and $\omega_c \geq 3.5$ Hz -- requiring a meaningful closed-loop bandwidth as done, \emph{e.g.}, in a similar problem in \cite{gimondi2021mixed}. The control tuning rule $\mathcal{R}\left(\mathcal{M}\left(\Theta\right)\right)$ is thus loop shaping, as implemented via the MATLAB function \verb|pidtune|.

\section{Simulation analysis}
\label{section:simulation_analysis}
In this section, we apply the neural-network approach described in Section \ref{section:control_design} to the case study of Section \ref{section:case_study}, showcasing its performance in terms of learning the control design policy.

\medskip

\noindent \textit{A. Dataset generation}

In order to train NNs, we generate a set of data from the yaw-rate dynamics model. We set $M=40000$, so as to obtain a meaningful number of examples; they are randomly split between $90\%$ for training and $10\%$ for validation. \\Further, we generate $M=1000$ more scenarios to evaluate NNs on a different dataset (called the \emph{testing set}). The testing set is also used to evaluate the benchmark identification and control tuning approaches, which we are going to describe in the following. The steer input $\delta_{f,k},\ k=1,\ldots,N$, is selected to be a Pseudo-Random-Binary signal, so as to provide a meaningful excitation to the system; the output yaw-rate $r_k,\ k=1,\ldots,N$, is perturbed with Gaussian distributed white noise, in order to achieve a signal-to-noise ratio of $\approx 30$ dB.

\medskip 

\noindent \textit{B. Gray-box identification benchmark}

As a benchmark to compare the neural-network approach against, we consider gray-box (GB) identification. This benchmark is possibly the best solution to be carried out when the model structure and some initial value of the parameters are approximately known, and only minor refinements are necessary. Given the initial model $\mathcal{M}\left(\Theta_0\right)$, we refine it using available data, obtaining $\mathcal{M}_{gb}$. The prediction error minimization approach can be utilized to adjust the model uncertain parameters (we consider here the MATLAB implementation \verb|pem|).\\
Once the refined model has been identified, one can apply the loop shaping tuning rule described in Section \ref{section:control_design}.\\

\noindent \textit{C. Black-box identification benchmark}

In case the exact model equations are not known, one has to rely on black-box (BB) identification. 
Taking the available input-output data as for the GB identification approach, a possible workflow is \cite{formentin2014comparison}:
\begin{enumerate}
\item Select the best auto-regressive with external input (ARX) model, by choosing the order yielding the lowest prediction loss;
\item use the found model order to train different polynomial models: specifically, we consider ARX, auto-regressive moving-average with external input (ARMAX), Box-Jenkins (BJ) and Output-Error (OE) models;
\item validate the models on a different dataset so as to select the best performing one here, rather than on the training set;
\item the best performing model $\mathcal{M}_{bb}$ is then used to tune a controller via the loop-shaping rule $\mathcal{R}$.
\end{enumerate}

\medskip

\noindent \textit{D. Virtual Reference Feedback Tuning}

A different philosophy when dealing with control design is the Virtual Reference Feedback Tuning (VRFT) \cite{campi2002virtual}. VRFT is a \textit{direct} method, that allows the user to obtain a parametric controller via input-output data, without going through an identification stage. However, unlike the Neural Network method proposed above, VRFT is not able to learn an \textit{a-priori} tuning rule $\mathcal{R}$: instead, it minimizes the difference between the closed-loop behaviour of the system under control and a reference model $M_r\left(z\right)$:
\begin{equation}
		\phi_{vrft}\left(\theta\right)=\argmin_{\phi}\left|\left| \dfrac{\mathcal{M}\left(z\right)C\left(\phi\right)}{1+\mathcal{M}\left(z\right) C \left(\phi\right)}-M_r\left(z\right) \right| \right|_2^2.
		\label{eq:vrft_cost}
\end{equation}
In order to compare VRFT to the other approaches, which are based on loop shaping, we select the reference model as a (discrete-time) low-pass filter with unitary gain and a single pole in $\omega_c$ -- the cutting frequency; in fact, resorting to basic control theory, if the phase margin $\varphi_m \geq 80^\circ$, we can safely approximate the closed-loop behaviour in this way.

\medskip 

\noindent \textit{E. CNN training details}

The considered time-series for training and validation are composed of $250$ samples and two channels, for input and output data respectively. Hence, the CNN input $\mathcal{I}$ has a size of $250 \times 2$.
The first convolution has a $2\times2$ kernel, with $96$ filters, allowing to capture short-term effects, while the following two capture the long-lasting effects, with kernels of size $120\times 1$ and $20\times 1$, and $48$ and $96$ filters, respectively. Indeed, the kernel size selection depends on the input size.
The number of filters have been found after some trial-and-error attempts.\\ We expect that the overall functioning of the network is equivalent to that of extracting features that mimic an identification procedure and the application of the control design rule.\\

\noindent \textit{F. WN training details}

The training and validation datasets for the WN consist of $500$ samples each, where each sample is a tuple of input and output data. A kernel size of $K=5$ is chosen and the number of layers is $L=10$. For each causal 1-D convolution performed in the WN, the number of output channels is chosen to be $4$. In the final 1-D convolution that gives $\hat{\phi}$, as depicted in Fig.~\ref{fig:Wavenet_unit}, the number of output channels is $1$. Since $k_p$ and $T_i$ are the controller parameters, we train two WNs in parallel, which output the estimates of $k_p$ and $T_i$ respectively.  Adam~\cite{kingma2014adam} is chosen as the optimizer to train WN with learning rate \verb|lr| $= 1e^{-3}$ and $\beta=(\beta_1, \beta_2) = (0.9, 0.999)$. The number of epochs is $50$.  

\begin{figure}[h]
\begin{subfigure}{0.24\textwidth}
\centering
	\includegraphics[width=0.9\linewidth]{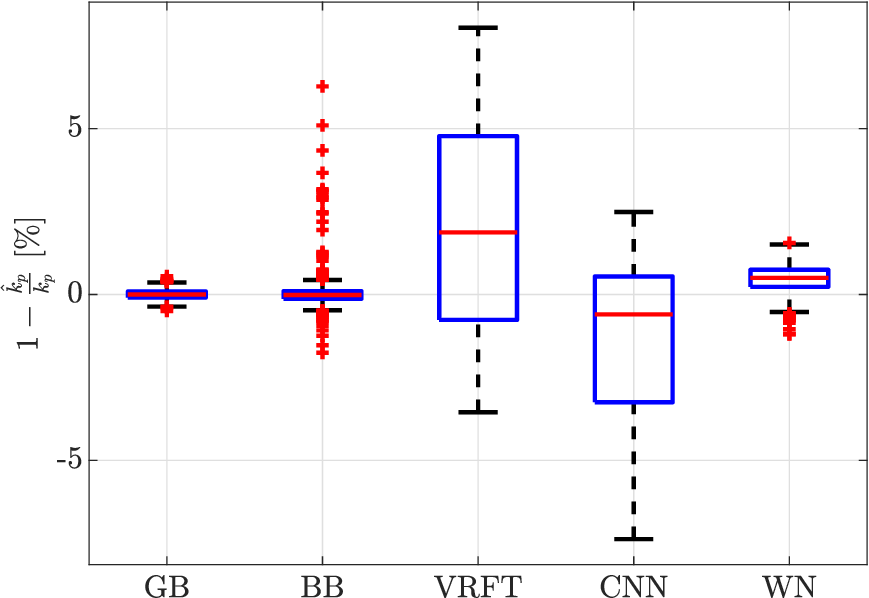}
	\caption{Box-plot of the errors in $k_p$.}
	\label{fig:boxplot_kp}
\end{subfigure}
\hfill 
\begin{subfigure}{0.24\textwidth}
	\centering
	\includegraphics[width=0.9\linewidth]{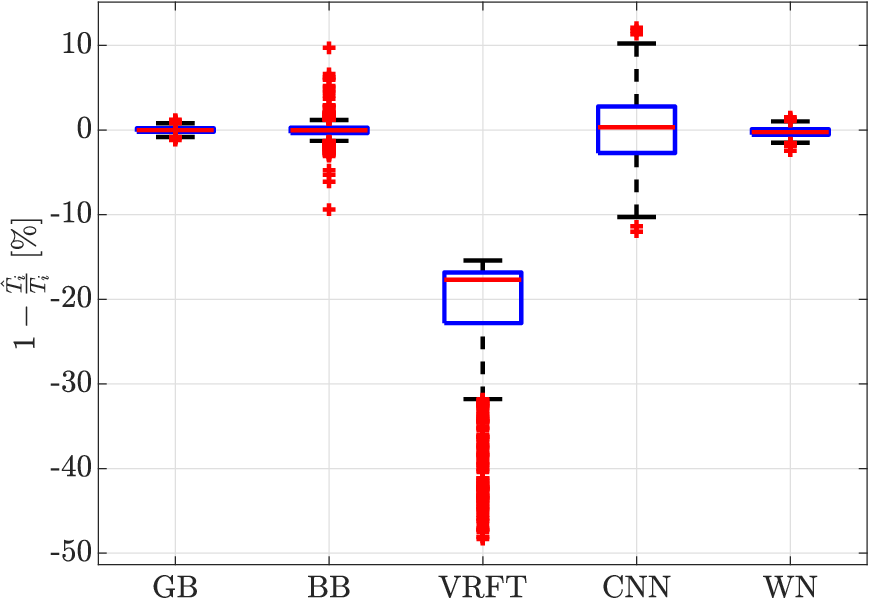}
	\caption{Box-plot of the errors in $T_i$.}
	\label{fig:boxplot_Ti}
\end{subfigure}


\label{Fig: Box plot and phase margin}
\caption{Box plot comparison.}
\end{figure}


\begin{table}[h]
\centering

\scalebox{0.7}{
\begin{tabular}{|l|c|c|}
\hline
     & $rms(1-k_p/\hat{k}_p)\times 100$ $[\%]$ & $rms(1-T_i/\hat{T_i}) \times 100$ $[\%]$ \\ \hline
GB   & $0.14$                                             & $0.33$                                              \\ \hline
BB   & $0.75$                                             & $1.01$      
                             \\ \hline
VRFT & $3.80$                                             & $22.79$                                             \\ \hline
CNN  & $2.57$                                             & $4.09$                                              \\ \hline
WN  & $0.63$                                             & $0.58$  
                             \\ \hline
\end{tabular}
}
\captionsetup[table]{skip=5pt} 
\caption{Controller parameters regression performance, in terms of relative error of the actual parameters.}
\label{tab:controller_learning_results}
\end{table}

\medskip

\noindent \textit{G. Controller parameters identification performance}

Fig.~\ref{fig:boxplot_kp} depicts the percentage error in estimating the controller proportional gain $k_p$, in the testing set. As one can note, the identification-based approaches achieve better results in terms of variance of the estimate. However, for all considered approaches, most of the errors are smaller than $5\%$, which is, as we will see further on, perfectly acceptable.
VRFT exhibits some bias, and this is expected, as the method works in a slightly different way (as we said, it does not learn the tuning rule $\mathcal{R}$). Similar results are obtained for the integral time $T_i$, see Fig.~\ref{fig:boxplot_Ti}; the bias in VRFT estimate is here even larger. 

\begin{figure}[h]
    \centering
	\includegraphics[width=0.8\columnwidth]{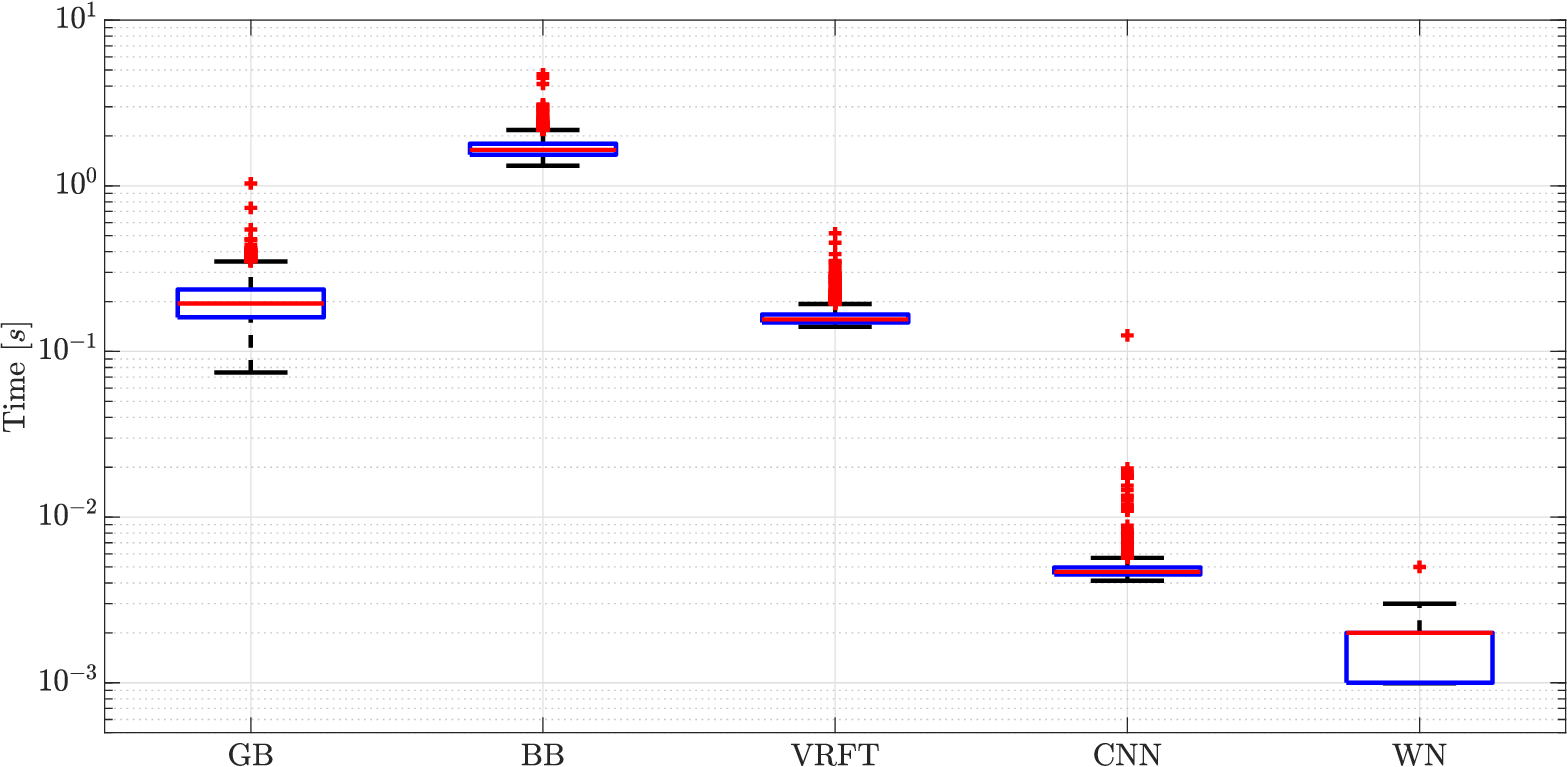}
	\caption{Computational time analysis for the different methods.}
 \label{fig:time_analysis}
\end{figure}

From Table~\ref{tab:controller_learning_results} one can see that our proposed neural-network based approaches attain low root mean square error (\emph{rms}) as compared to BB and VRFT. Of the proposed neural-network approaches, WN attains remarkable accuracy (low \emph{rms}) in the estimation of the controller parameters; this is due to the fact that
WN captures long-term causal dependencies in the input-output data, and the operations in its dilated layers serve as a ``memory" similar to the state in LSTMs~\cite{HochSchm97}. 
Fig.~\ref{fig:time_analysis} shows boxplots of the computational times for each method. It is clear that our proposed methods are significantly faster than BB and GB. In particular, WN is faster than CNN and this is due to the fact that WNs require less operations compared to CNNs.


Now, it is interesting to assess what the results in Fig.~\ref{fig:boxplot_kp} mean: how does an error in estimating the controller parameters converts into the phase margin requirement?
\begin{figure}[h]
\centering
\includegraphics[width=0.6\linewidth]{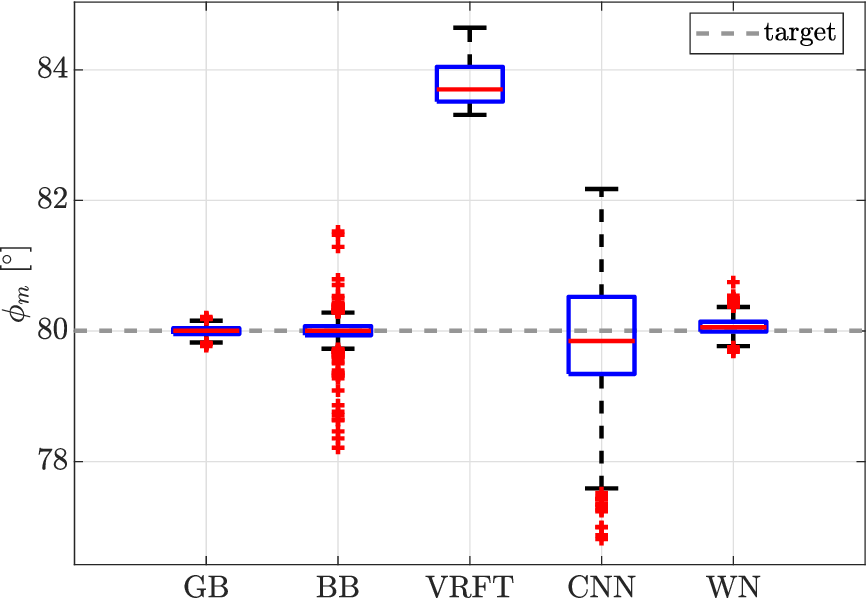}
	\caption{Phase margin comparison.}
	\label{fig:phase_margin}
\end{figure}

\medskip

\noindent \textit{H. Closed-loop performance}

Fig.~\ref{fig:phase_margin} shows the actual phase margin obtained using the values of $k_p$ and $T_i$ estimated by the different methods. As we can note, all methods guarantee acceptable values for both variables: all controllers are stable, and no practical differences exist between a phase margin of $78^\circ$ and one of $80^\circ$ -- as can be seen, for example, by looking at some ``worst cases'' for the CNN and WN predictions. \\
\section{Conclusion} \label{sec:conclusion}
In this paper, we showed how a control design rule can be learned by means of neural networks. Leveraging the use of a Digital Twin, we generate large amounts of artificial data. For each batch of data, we design a controller. These data-controller pairs are then used to train a black-box model via supervised learning; the model is then directly used on fresh data for finding the controller satisfying the learned rule. \\
Many future developments indeed open up; it would be interesting to test this methodology on different applications, possibly obtaining an estimator not application-dependent, but rather tuning-rule-dependent. Also, an important research direction consists in providing formal guarantees for the presented methodology.

\bibliographystyle{IEEEtran}
\bibliography{refs_IFAC}             

\begin{thebibliography}{10}
\providecommand{\url}[1]{#1}
\csname url@samestyle\endcsname
\providecommand{\newblock}{\relax}
\providecommand{\bibinfo}[2]{#2}
\providecommand{\BIBentrySTDinterwordspacing}{\spaceskip=0pt\relax}
\providecommand{\BIBentryALTinterwordstretchfactor}{4}
\providecommand{\BIBentryALTinterwordspacing}{\spaceskip=\fontdimen2\font plus
\BIBentryALTinterwordstretchfactor\fontdimen3\font minus
  \fontdimen4\font\relax}
\providecommand{\BIBforeignlanguage}[2]{{%
\expandafter\ifx\csname l@#1\endcsname\relax
\typeout{** WARNING: IEEEtran.bst: No hyphenation pattern has been}%
\typeout{** loaded for the language `#1'. Using the pattern for}%
\typeout{** the default language instead.}%
\else
\language=\csname l@#1\endcsname
\fi
#2}}
\providecommand{\BIBdecl}{\relax}
\BIBdecl

\bibitem{aastrom2006advanced}
K.~J. {\AA}str{\"o}m and T.~H{\"a}gglund, \emph{Advanced PID control}.\hskip
  1em plus 0.5em minus 0.4em\relax ISA-The Instrumentation, Systems and
  Automation Society, 2006.

\bibitem{samad2017survey}
T.~Samad, ``A survey on industry impact and challenges thereof [technical
  activities],'' \emph{IEEE Control Systems Magazine}, vol.~37, no.~1, pp.
  17--18, 2017.

\bibitem{tanelli2011transmission}
M.~Tanelli, G.~Panzani, S.~M. Savaresi, and C.~Pirola, ``Transmission control
  for power-shift agricultural tractors: Design and end-of-line automatic
  tuning,'' \emph{Mechatronics}, vol.~21, no.~1, pp. 285--297, 2011.

\bibitem{dettu2023til}
F.~Dettù, S.~Formentin, and S.~M. Savaresi, ``The twin-in-the-loop approach
  for vehicle dynamics control,'' \emph{IEEE/ASME Transactions on
  Mechatronics}, pp. 1--12, 2023.

\bibitem{liu2021review}
M.~Liu, S.~Fang, H.~Dong, and C.~Xu, ``Review of digital twin about concepts,
  technologies, and industrial applications,'' \emph{Journal of Manufacturing
  Systems}, vol.~58, pp. 346--361, 2021.

\bibitem{ljung1999system}
L.~Ljung, \emph{System Identification: Theory for the User}, 2nd~ed.\hskip 1em
  plus 0.5em minus 0.4em\relax Prentice Hall, 1999.

\bibitem{piga2017direct}
D.~Piga, S.~Formentin, and A.~Bemporad, ``Direct data-driven control of
  constrained systems,'' \emph{IEEE Transactions on Control Systems
  Technology}, vol.~26, no.~4, pp. 1422--1429, 2017.

\bibitem{formentin2012non}
S.~Formentin, S.~Savaresi, and L.~Del~Re, ``Non-iterative direct data-driven
  controller tuning for multivariable systems: theory and application,''
  \emph{IET control theory \& applications}, vol.~6, no.~9, pp. 1250--1257,
  2012.

\bibitem{formentin2014comparison}
S.~Formentin, K.~Van~Heusden, and A.~Karimi, ``A comparison of model-based and
  data-driven controller tuning,'' \emph{International Journal of Adaptive
  Control and Signal Processing}, vol.~28, no.~10, pp. 882--897, 2014.

\bibitem{qi2019tuning}
Z.~Qi, Q.~Shi, and H.~Zhang, ``Tuning of digital {PID} controllers using
  particle swarm optimization algorithm for a can-based dc motor subject to
  stochastic delays,'' \emph{IEEE Transactions on Industrial Electronics},
  vol.~67, no.~7, pp. 5637--5646, 2019.

\bibitem{khosravi2021performance}
M.~Khosravi, V.~N. Behrunani, P.~Myszkorowski, R.~S. Smith, A.~Rupenyan, and
  J.~Lygeros, ``Performance-driven cascade controller tuning with {Bayesian}
  optimization,'' \emph{IEEE Transactions on Industrial Electronics}, vol.~69,
  no.~1, pp. 1032--1042, 2021.

\bibitem{lucchini2020torque}
A.~Lucchini, S.~Formentin, M.~Corno, D.~Piga, and S.~M. Savaresi, ``Torque
  vectoring for high-performance electric vehicles: an efficient mpc
  calibration,'' \emph{IEEE Control Systems Letters}, vol.~4, no.~3, pp.
  725--730, 2020.

\bibitem{linaro2023continuous}
D.~Linaro, F.~Bizzarri, D.~Del~Giudice, C.~Pisani, G.~M. Giannuzzi, S.~Grillo,
  and A.~M. Brambilla, ``Continuous estimation of power system inertia using
  convolutional neural networks,'' \emph{Nature Communications}, vol.~14,
  no.~1, p. 4440, 2023.

\bibitem{karlsson2021speed}
R.~Karlsson and G.~Hendeby, ``Speed estimation from vibrations using a deep
  learning {CNN} approach,'' \emph{IEEE Sensors Letters}, vol.~5, no.~3, pp.
  1--4, 2021.

\bibitem{wu2017introduction}
J.~Wu, ``Introduction to convolutional neural networks,'' \emph{National Key
  Lab for Novel Software Technology. Nanjing University. China}, vol.~5,
  no.~23, p. 495, 2017.

\bibitem{oord2016wavenet}
A.~v.~d. Oord, S.~Dieleman, H.~Zen, K.~Simonyan, O.~Vinyals, A.~Graves,
  N.~Kalchbrenner, A.~Senior, and K.~Kavukcuoglu, ``Wavenet: A generative model
  for raw audio,'' \emph{arXiv preprint arXiv:1609.03499}, 2016.

\bibitem{beal2012model}
C.~E. Beal and J.~C. Gerdes, ``Model predictive control for vehicle
  stabilization at the limits of handling,'' \emph{IEEE Transactions on Control
  Systems Technology}, vol.~21, no.~4, pp. 1258--1269, 2012.

\bibitem{sierra2006cornering}
C.~Sierra, E.~Tseng, A.~Jain, and H.~Peng, ``Cornering stiffness estimation
  based on vehicle lateral dynamics,'' \emph{Vehicle System Dynamics}, vol.~44,
  no. sup1, pp. 24--38, 2006.

\bibitem{corno2021optimization}
M.~Corno, A.~Gimondi, G.~Panzani, F.~Roselli, A.~Alessandretti, and S.~M.
  Savaresi, ``A non-optimization-based dynamic path planning for autonomous
  obstacle avoidance,'' \emph{IEEE Transactions on Control Systems Technology},
  vol.~31, no.~2, pp. 722--734, 2023.

\bibitem{gimondi2021mixed}
A.~Gimondi, M.~Corno, and S.~M. Savaresi, ``A mixed sideslip yaw rate stability
  controller for over-actuated vehicles,'' in \emph{International Design
  Engineering Technical Conferences and Computers and Information in
  Engineering Conference}, vol. 85369, 2021, p. V001T01A001.

\bibitem{campi2002virtual}
M.~C. Campi, A.~Lecchini, and S.~M. Savaresi, ``Virtual reference feedback
  tuning: a direct method for the design of feedback controllers,''
  \emph{Automatica}, vol.~38, no.~8, pp. 1337--1346, 2002.

\bibitem{kingma2014adam}
D.~P. Kingma and J.~Ba, ``Adam: A method for stochastic optimization,''
  \emph{arXiv preprint arXiv:1412.6980}, 2014.

\bibitem{HochSchm97}
S.~Hochreiter and J.~Schmidhuber, ``Long short-term memory,'' \emph{Neural
  Computation}, vol.~9, no.~8, pp. 1735--1780, 1997.

\end{thebibliography}
	
\end{document}